\documentclass[12pt]{article}
\usepackage{graphicx,amsmath}
\usepackage{color}

\newlength{\dinwidth}
\newlength{\dinmargin}
\def\be{\begin{equation}}   
\def\ee{\end{equation}}  
\def\bea{\begin{eqnarray}}                      
\def\eea{\end{eqnarray}}
\def\ch1{$\chi(1^+)$}
\def\lapproxeq{\lower .7ex\hbox{$\;\stackrel{\textstyle                                                    
<}{\sim}\;$}}                                                    
\def\gapproxeq{\lower .7ex\hbox{$\;\stackrel{\textstyle                                                    
>}{\sim}\;$}}

\begin{document}

\begin{flushright}                                                    
IPPP/17/35 \\
\today \\                                                    
\end{flushright} 

\vspace*{0.5cm}

\vspace*{0.5cm}
\begin{center}
{\Large \bf Total $\pi^+ p$ cross section extracted from the} \\ 
\vspace{0.2cm}
{\Large \bf leading neutron spectra at the LHC}\\
\vspace*{1cm}

V.A. Khoze$^{a,b}$, A.D. Martin$^{a}$ and M.G. Ryskin$^{a,b}$\\

\vspace*{0.5cm}
$^a$ Institute for Particle Physics Phenomenology, Durham University, 
Durham, DH1 3LE, UK\\ 

$^b$ Petersburg Nuclear Physics Institute, NRC `Kurchatov Institute', 
Gatchina, St.~Petersburg, 188300, Russia \\
\end{center}

\begin{abstract}
We use the very forward neutron energy spectra measured by the Large Hadron Collider forward (LHCf) experiment at 7 TeV to extract the $\pi^+ p$ total cross section at centre-of-mass energies in the range 2.3$-$3.5 TeV.
To do this we have to first isolate the $\pi$-exchange pole in forward neutron production in $pp$ collisions, by evaluating other possible contributions. Namely, those from $\rho$ and $a_2$ exchange, from both eikonal and enhanced screening effects, from migration, from neutron production by $\Delta$-isobar decay and from diffractive nucleon excitations. We discuss the possible theoretical uncertainties due to the fact that the data do not exactly reach the $\pi$  pole. We choose the kinematical domain where the pion contribution dominates and demonstrate the role of the different corrections which could affect the final result.

\end{abstract}
\vspace*{0.5cm}

\section{Introduction}
The recent LHCf measurements of leading neutron production at 7 TeV~\cite{LHCf}
 have boosted the interest in attempts to extract 
 the high energy
 pion-proton cross section from these data, see for instance \cite{Petrov:2009wr,Kopeliovich:2014wxa,Ryutin:2016hyi}.
This, in turn, would allow new discriminative  tests of the existing models of high-energy
hadron interactions. Recall that at present the results of direct measurements
of the $\pi^+ p$ cross sections are known only up to $\sqrt s=25$ GeV \cite{PDG}.
In order to extend the pion-proton interaction energy range various 
indirect methods for  extraction of the $\pi p$ cross section
were proposed in the literature, see for example \cite{Soding:1965nh,Ryskin:1997zz,
Breitweg:1997ed,Sobol:2010mu}. All these approaches are based one way or another on the assumption
that one can reliably isolate the  pion-exchange contribution in the corresponding
processes. This topic has a long and chequered history (see e.g. 
\cite{Chew:1958wd}
$-$ \cite{Kop2}).

The idea of using the inclusive leading neutron spectra in high-energy
proton collisions for the separation of the pion-exchange contribution is to exploit
the natural conjecture that due to the small value of the pion mass this term should
play an important, or even, dominant role.
The position of the pion pole is rather close to the physical region and {\em if} it were possible to measure the cross section just at the pole then undoubtedly we would deal with pure pion exchange.
In particular, in such a case the absorptive corrections,
caused by rescattering effects
 (see, for example, \cite{Nikolaev:1997cn,DAlesio:1998uav,Kaidalov:2006cw})
 would be negligible, 
and the value of the so-called  survival factor, $S^2$, of the rapidity gap associated with $\pi$ -exchange
would be close to 1, $S^2=1$. 

The problem is that we cannot reach the pole,
which is outside the physical region, and the only way is 
to focus on a limited kinematic domain, located close to the $\pi$-pole,
and then to evaluate the size of the various corrections caused by the extrapolation
to $m_{\pi}^2$.
 The main effects are:
 \begin{itemize}
\item[(i)]the contributions from the $\rho$ and $a_2$ Regge trajectories which have intercepts
 higher than that for the pion; these terms will dominate when the momentum 
fraction carried by the leading neutron $x_L\to 1$,
\item[(ii)] absorptive corrections, that is a gap survival factor $S^2<1$,
\item[(iii)] leading neutrons produced in the decays of higher proton excitations such as
 $N^*(1440)$ or the $\Delta$ isobar,
\item[(iv)] migration~\cite{Kaidalov:2006cw} of the leading neutron due to baryon rescattering.
\end{itemize}

In Section 2 we recall the expressions for the inclusive neutron cross section caused by the pion and the secondary Reggeon exchanges, then in Section 3 we consider the screening (or absorptive) corrections. In Section 4 we consider in detail those kinematic domains of the LHCf forward neutron data \cite{LHCf} which allow us to sufficiently isolate the $\pi$-exchange contribution so as to obtain reliable values of the $\pi p$ total cross section. We find that
to be closer to the pion pole and to minimize the transverse momentum effects we should choose 
LHCf data from the largest rapidity interval ($\eta>10.76$) and to concentrate on the three bins of the
 neutron energy $E_n=3.25-3,\ 3-2.75,\ 2.75-2.5$ TeV. In the largest $x_L$ bin ($E_n=3.5-3.25$ TeV) the experimental error and the possible contribution from $\rho$ and $a_2$ trajectories are too large, while at lower $x_L$ values the pion has larger virtuality due to the longitudinal component of its momentum ($t_{min}=(1-x_L)^2m^2_N/x_L$), and the contribution from baryon rescattering, that is from migration, becomes non-negligible.

We present our results for $\sigma^{\rm tot}(\pi p)$ in Section 5.  In Section 6 we use the same formalism to describe the old lower-energy CERN-ISR data.  Our conclusions and the outlook for applying the formalism to future leading neutron data are presented in Sections 7 and 8 respectively.

 \section{Born-level cross sections}
 In this section we evaluate the contributions to the cross section for forward neutron production in $pp$ collisions coming from $\pi,~\rho$ and $a_2$ exchanges, and from baryon excitations of the protons.

\subsection{Pion exchange}
Neglecting absorptive effects, the contribution of reggeized pion exchange to the inclusive neutron production reads
\be
\label{pi-reg}
\frac{x_Ld\sigma^\pi (pp\to nX)}{dx_Ldq^2_t}
=\frac{G^2_{\pi^+pn}(-t)}{16\pi^2(t-m^2_\pi)^2}
F^2_{\pi N}(t)\sigma^{\rm tot}_{\pi p}(M^2_X)(1-x_L)^{1-2\alpha_\pi(t)}\ ,
\ee
where $\alpha_\pi(t)=\alpha'_\pi(t-m^2_\pi)$ is the pion trajectory
with slope
$\alpha'_\pi=0.9$ GeV$^{-2}$ 
 and  coupling 
$G^2_{\pi^+ pn}/8\pi=13.75$ ~\cite{Stoks:1992ja,Arndt:1995bj}. 
The formulae for the invariant mass $M_X$ of the produced system $X$ and of $-t$ are  given
 by 
 \be M^2_X=s(1-x_L),
 \ee
 \be 
 -t=(1-x_L)^2m_N^2/x_L+q^2_t/x_L~,
 \ee
  where $q_t$ is the neutron 
transverse momentum and $m_N$ is the nucleon mass.
 
 Here we have retained in the reggeon signature factor $\eta_\pi(t)$ only the denominator $1/(t-m^2_\pi)$ while the remaining $t$ dependence is absorbed in the effective vertex form factor $F_{\pi N}(t)$.
Below we will use the non-reggeized version of (\ref{pi-reg})
\be
\label{pi-noreg}
\frac{x_Ld\sigma^\pi (pp\to nX)}{dx_Ldq^2_t}
=\frac{G^2_{\pi^+pn}(-t)}{16\pi^2(t-m^2_\pi)^2}
F^2_{\pi N}(t)\sigma^{\rm tot}_{\pi p}(M^2_X)(1-x_L)\ ,
\ee
 with a dipole parametrization of the form factor
\be
F_{\pi N}(t)=1/(1+(m^2_\pi-t)/0.71\mbox{GeV}^2)^2.  
\ee
In such a form, (\ref{pi-noreg}), the interpretation of the result in terms of the $\pi p$ cross section is more straightforward.
It is possible to slightly modify the expression for $F_{\pi N}(t)$. This does not change the result noticeably. Moreover,
since we work in the small $|t|$-domain, where the pion trajectory 
$\alpha_\pi(t)$ is close to zero, in both the reggeized and non-reggeized cases, we get practically the same result.

\subsection{Secondary trajectories}
Another contribution to the leading neutron spectrum
is generated by
the exchange of $\rho$ and $a_2$ isovector trajectories. Due to their larger 
intercepts $\alpha_{\rho,a_2}(0)\simeq 0.5$, this contribution should dominate as $x_L\to 1$.
We write the cross section arising for $\rho$ exchange in a form analogous to (\ref{pi-reg})

\be
\label{rho-reg}
\frac{x_Ld\sigma^\rho (pp\to nX)}{dx_Ldq^2_t}=|\eta(t)|^2
\frac{g^2_{\rm nf}+g^2_{\rm sf}q^2_t/4m^2_N}{16\pi^2(t-m^2_\rho)^2}
F^2_{\rho N}(t)\sigma^{\rm tot}_{\rho p}(M^2_X)(1-x_L)^{1-2\alpha_\rho(t)}\ .
\ee
We assume `exchange degeneracy' (see, for example,~\cite{Collins}) between the $\rho$ and $a_2$ exchanges. That is, the trajectory $\alpha_{a_2}(t)=\alpha_\rho(t)=0.54+\alpha't$ (with $\alpha'=0.9$ GeV$^{-2}$). Moreover, this means that the $\rho$ and $a_2$ trajectories have the same 
residues and vertex form factors. The only difference is the signature factor
\be
\eta(t)=\frac 12[1\pm\exp(-i\pi\alpha_R(t))]
\ee
with a plus sign for $a_2$-exchange and a minus sign for $\rho$-exchange ($R=\rho, a_2$). This means that when  $a_2$-exchange is included, we have to replace
  the first factor $|\eta(t)|^2$ in ({\ref{rho-reg}) by 1.
\be
\label{rho-a2-reg}
\frac{x_Ld\sigma^{\rho+a_2} (pp\to nX)}{dx_Ldq^2t}=
\frac{g^2_{\rm nf}+g^2_{\rm sf}q^2_t/4m^2_N}{16\pi^2(t-m^2_\rho)^2}
F^2_{\rho N}(t)\sigma^{\rm tot}_{\rho p}(M^2_X)(1-x_L)^{1-2\alpha_\rho(t)}\ .
\ee
Here $g_{\rm sf}$ and $g_{\rm nf}$ are the couplings corresponding to the processes where
the neutron helicity is opposite (spin flip) to that of the incoming proton or the same (non flip) as the proton helicity~\footnote{Recall that in the case of pion exchange we also have `flip' and `non flip' contributions hidden in the factor $-t=q^2_t+q^2_L=q^2_t+t_{min}+(1-x_L)q^2_t/x_L$,
 where the first term, $q^2_t$, corresponds to spin flip
production while the second term, $q^2_L$, describes the non-flip process.}
 and we use $F_{\rho N}(t)=$ exp$(B_\rho t)$ with 
 $B_\rho=2.3$ GeV$^{-2}$.~\footnote{This value is consistent with the slope observed for 
 the RRP term in the triple-Regge analysis~\cite{RRP}, accounting for the fact that part of 
 the RRP slope comes from the reggeon trajectory term $\alpha'_R\ln(1/(1-x_L))$. Here R=$\rho,~a_2$ reggeons and P=pomeron.}
 
Contrary to the pion-proton coupling $G$, which is known to rather good accuracy, there
are no accepted values for the $\rho\ (a_2)$-nucleon vertices.
The couplings $g_{\rm nf}$ and $g_{\rm sf}$ can be obtained from old Regge phenomenology,
say, from~\cite{Irving}, or alternately can be based on the Vector Meson Dominance (VMD) 
 model~\cite {VMD}. Since absorptive corrections were not accounted for in the old Regge phenomenological description we prefer to use the VMD-based values which are larger (see Table AA3 of \cite{Irving}). That is, our estimate of the correction caused by the $\rho,\ a_2$ contribution may be considered as `conservative'. Finally, following
the additive quark model we assume that $\sigma(\pi p)=\sigma(\rho p)=\sigma(a_2 p)$.

\subsection{Baryon excitations}
For $\Delta$-isobar production via pion and $\rho,\ a_2$ exchanges we use formulae analogous to (\ref{pi-reg},\ref{rho-a2-reg}) with couplings taken from Table AA3 of~\cite{Irving}. Since the different helicity states of the $\Delta$ are produced with different couplings we account for the polarization effects in $\Delta\to n\pi$ decay.

Larger uncertainties may result from neutrons coming from the decays of the $N^*$
resonances produced via diffractive proton dissociation. Currently there are no 7 TeV data on the cross section and polarization of the corresponding resonances. 
The only more or less relevant experimental cross section is the TOTEM
result for low mass ($M_X<3.4$ GeV) proton dissociation,  $\sigma^D_{lowM}=2.62\pm 2.17$ mb~\cite{TOTEM1}. Proton excitations with $M_X>2$ GeV have a small probability to create a neutron with  large $x_L$ and small $q_T$; these states are decaying mainly  into multiparticle systems with two or more pions.
The main danger represents the contribution from 
the $M_X=1.3~-~1.8$ GeV region. In our computations we assume a non-polarized (isotropic) decay with the branching ratio Br$(p^*\to n\pi^+)\simeq 1/3$ and the corresponding cross section of
 one~proton excitation\footnote{Note that 2.6 mb corresponds to excitations of {\bf both} of the initial protons.} to be 1 mb, see Section \ref{sec:4.5}.

\section{Screening corrections}

Absorptive effects play an important role in processes where one particle carries away almost all of the beam energy; that is, its $x_L$ is close to 1. This leads to the formation of a rapidity gap, since the remaining energy is not large enough to produce secondaries in the forward rapidity interval. However, any interaction of the fast particle will decrease the value of its $x_L$ and thus diminish the cross section at large $x_L$.
For example, these absorptive or screening corrections were responsible for the breaking of factorization, by about an order of magnitude, in diffractive dijet production at the Tevatron \cite{KKMR2}.

There are two types of absorptive corrections.  These corrections are discussed in some detail in \cite{Kaidalov:2006cw}.  
First, we have the effects caused by the inelastic interactions between the fast incoming proton (or leading neutron) and the target proton. The secondary particles from these interactions populate the rapidity gap  separating  the neutron from the other hadrons, and carry away
energy from the leading neutron.
The corresponding correction is described by additional eikonal-like 
Pomeron exchanges and we denote it as the `eikonal' gap survival factor $S^2_{\rm eik}$. The corresponding diagrams are sketched in Fig.~\ref{fig:4}.
\begin{figure}
\begin{center}
\vspace*{-2.0cm}
\includegraphics[height=5cm]{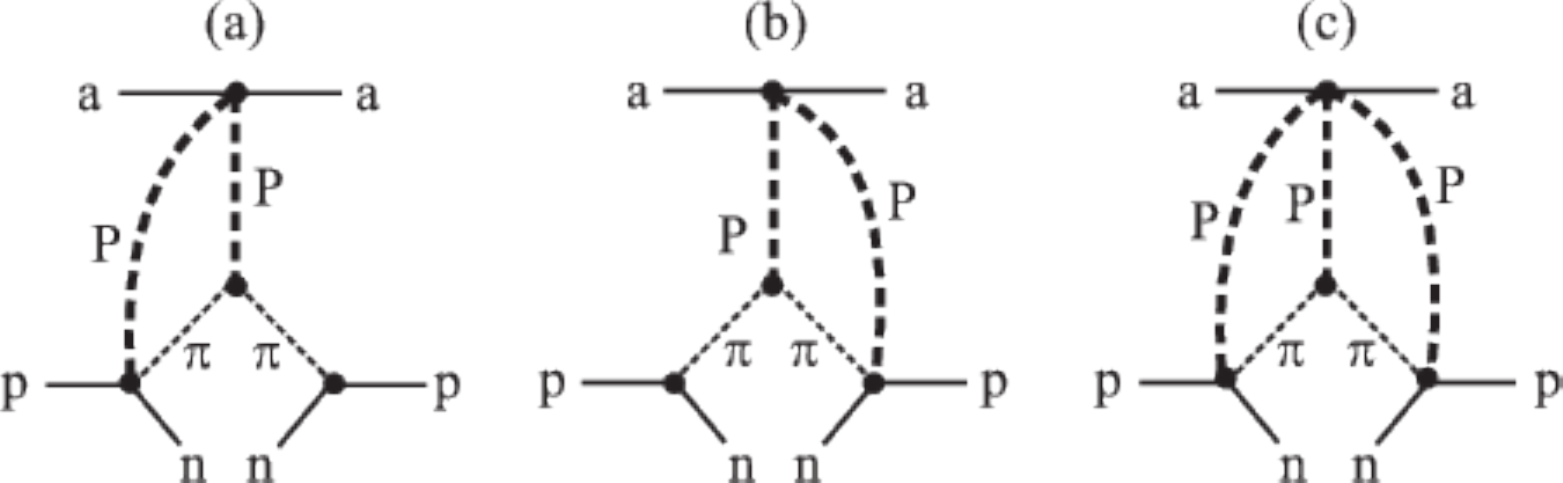}
\caption{\sf Symbolic diagrams of the eikonal absorptive corrections to the cross
section for the inclusive process $ap \to Xn$. In this paper hadron $a$ is a proton $p$, but in general the target particle $a$ can be any hadron.  The extra lines denoted by P,
which surround the triple-Regge interaction, represent multi-Pomeron exchanges
between the leading hadrons.
\label{fig:4}}
\end{center}
\end{figure}

\begin{figure}
\begin{center}
\vspace*{-0.5cm}
\includegraphics[height=5cm]{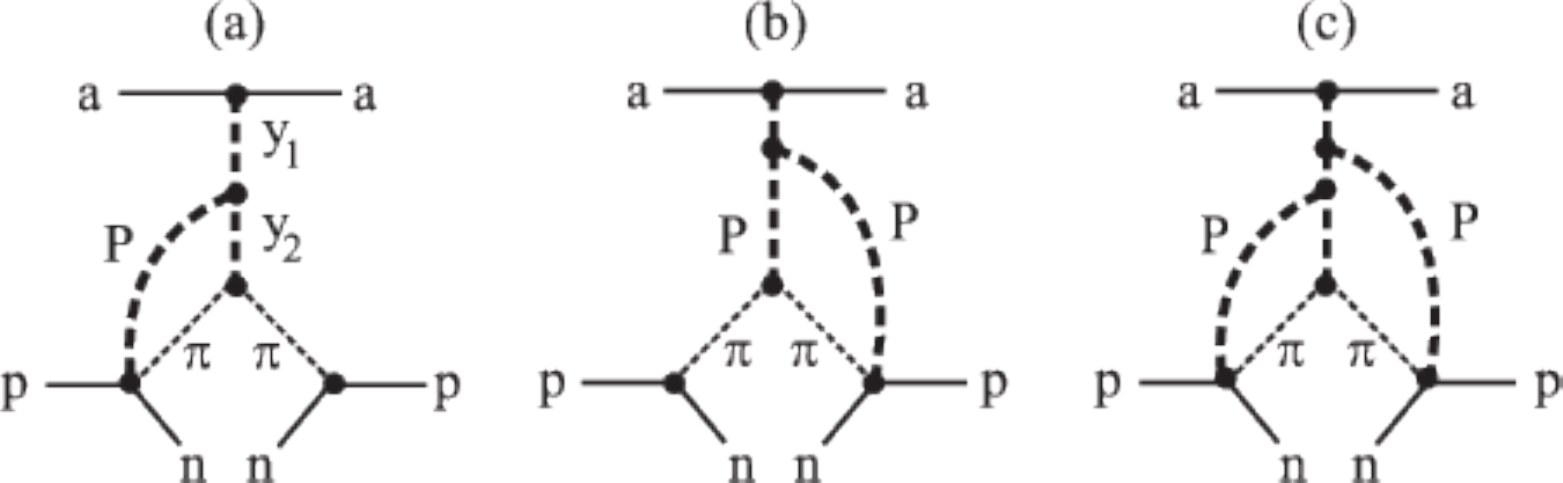}
\caption{\sf Symbolic diagrams for the ``enhanced" absorptive corrections to the cross
section for the inclusive process $ap \to Xn$, which become 
important at very high energies.    The extra lines denoted by P,
which are coupled directly to the ingoing $p$ or outgoing $n$,  represent multi-Pomeron exchanges.
\label{fig:5}}
\end{center}
\end{figure}

\begin{figure}
\begin{center}
\includegraphics[height=5cm]{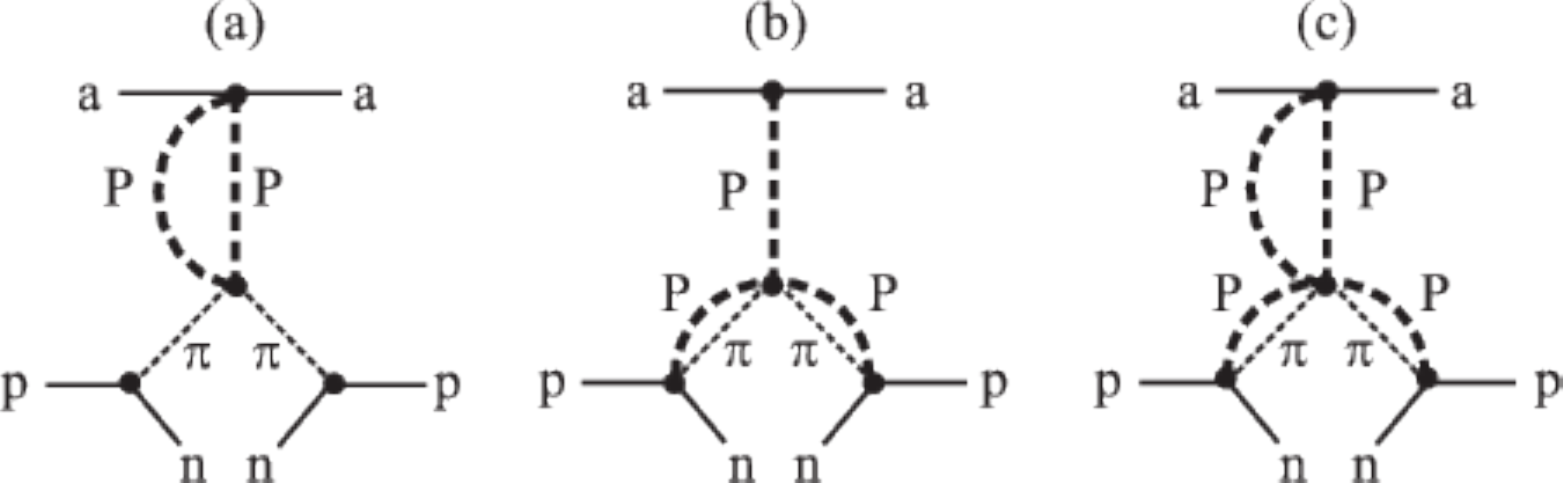}
\caption{\sf Multi-pomeron corrections to the reggeons in the triple-regge diagrams. 
\label{fig:6}}
\end{center}
\end{figure}
  Besides this, we have to consider an interaction 
(shown symbolically in Fig.~\ref{fig:5}) of the fast nucleon (proton or neutron) with the particles {\em within}
 the pion-target proton (or $\rho$-, $a_2$-proton) amplitude, that is in the remaining $X$ system.  This contribution could be 'enhanced' due to a large multiplicity of 
particles in this system $X$.
Therefore we denote the corresponding damping factor as $S^2_{\rm enh}$. The diagrams where an additional Pomeron screens the pion ($\rho,\ a_2$) propagator
 (such as shown in Fig.~\ref{fig:6}(b)) are also included in the $S^2_{\rm enh}$ factor, while the diagrams which describe an  interaction of the pion
 (or $\rho, \ a_2$) with the system $X$ (such as shown in Fig.~\ref{fig:6}(a)) are included in the $\pi$-proton ($\rho$-$,\ a_2$-proton) cross section.

\vspace*{-0.5cm}

\subsection{Eikonal survival factor}
\begin{figure}
\begin{center}
\includegraphics[height=10cm]{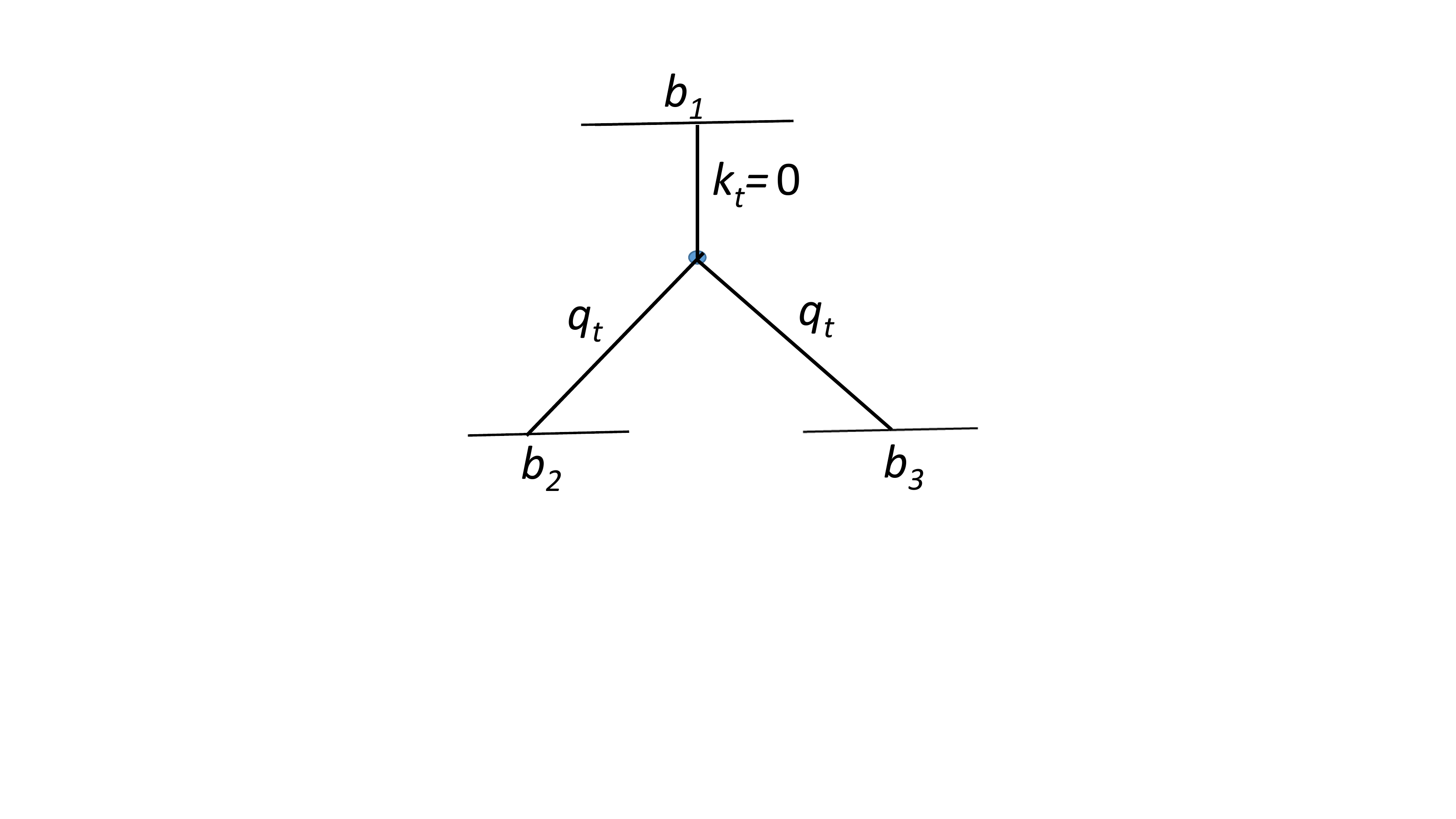}
\vspace*{-4.5cm}
\caption{\sf The conjugate variables used in (\ref{born-bt}).}
\label{fig:conj}
\end{center}
\end{figure}
To evaluate the most important (eikonal) screening correction we follow the approach of~\cite{luna} and
work in impact parameter, $b$ space. The `Born' cross section (\ref{pi-reg})
 can be written as (see Fig.~\ref{fig:conj} for the definition of the variables)
\be
\label{born-bt}
\frac{x_Ld\sigma^\pi}{dx_Ldq^2_t}=A\int\frac{d^2b_2}{2\pi}
e^{i\vec q_t\cdot\vec b_2}F^\pi(b_2)\int\frac{d^2b_3}{2\pi}
e^{i\vec q_t\cdot\vec b_3}F^\pi(b_3)\int\frac{d^2b_1}{2\pi}F^\sigma(b_1)\ ,
\ee 
where all factors which do not depend on the
transverse momentum are incorporated in the first factor $A$. The amplitudes
 $F^\pi(b_{2,3})$ are the Fourier conjugates of the pion exchange amplitudes written in $q_t$ space..
In particular, the spin non-flip amplitude reads 
\be
\label{F-pi}
F^\pi_{\rm nf}(b)=\int\frac{d^2q_t}{2\pi}e^{-i\vec q_t\cdot\vec b}~\frac{q_L F_{\pi N}(t)}
{t-m^2_\pi}(1-x_L)^{-\alpha_\pi(t)}\ ,
\ee
where $q_L$ is given by
\be
q_L=\sqrt{-t_{\parallel}}=\sqrt{(m^2_N(1-x_L)^2+(1-x_L)q^2_t)/x_L  }~,
\ee
and $m_N$ is the mass of the nucleon. Care should be taken in the case of a spin-flip amplitude since
 it depends on the {\em direction} of the transverse momentum $\vec q_t$. 
In practical terms this means that the angular integration results not
in a zero-order Bessel function $J_0(b_iq_t)$ but in the first order function $J_1(b_iq_t)\,$ (with $i=2,3)$.

 In the last integral $F^\sigma(b)$ corresponds to the pion-proton amplitude. To calculate this amplitude we use the same Pomeron-proton vertex form factor $F_{\rm Pom}$ and the same Pomeron trajectory slope, $\alpha'_{\rm Pom}$ as the ones  used in the model~\cite{1312} which
allows a good description of
the elastic proton-proton cross section measured by TOTEM~\cite{TOTEM2} at $\sqrt s=7$ TeV.
\be
\label{F-pom}
F^\sigma(b)=\int\frac{d^2k_t}{2\pi}e^{-i\vec k_t\cdot\vec b}
~F_{N-{\rm Pom}}(-k^2_t)~F_{\pi-{\rm Pom}}(-k^2_t)~x_L^{-k^2_t\alpha'_{\rm Pom}}
\ee
where the pion-Pomeron vertex form factor $F_{\pi-{\rm Pom}}(t)=\exp(B_\pi t)$ is parametrized by an exponent with  slope $B_\pi=2$ GeV$^{-2}$~\cite{Dino}. 
Note that there is no exponent $e^{i\vec k_t\cdot\vec b}$ in the last integral of (\ref{born-bt}) since this pion-proton amplitude is taken at $k_t=0$.

Now, in the $b$ representation, to account for the eikonal absorptive correction we have just to multiply the integrand of (\ref{born-bt}) by the screening factors
\be
\label{screen}
S_{\rm eik}(\vec b_2-\vec b_1)S_{\rm eik}(\vec b_3-\vec b_1)=
\exp(-\Omega(\vec b_2-\vec b_1)/2)\exp(-\Omega(\vec b_3-\vec b_1)/2)\ ,
\ee
where the proton-proton opacity $\Omega$ is taken from the model of~\cite{1312} which  reproduces well the elastic $pp$-cross section at 7 TeV.
That is, we have to compute the integral
\be
\label{Ib}
I^\pi(b_1)=\int\frac{d^2b}{2\pi}F^\pi(b)\exp(-\Omega(\vec b-\vec b_1)/2)
\ee
and to write the cross section as
\be
\frac{x_Ld\sigma^\pi}{dx_Ldq^2_t}=A\int\frac{d^2b_1}{2\pi}F^\sigma(b_1)~
|I^\pi(b_1)|^2\ .
\ee
For the exchange of the $\rho$ and $ a_2$ trajectories the gap survival factor $S^2_{\rm eik}$ is accounted for in a similar way.

 Up to now we described the calculation within the
framework of a single-channel eikonal model
 which does not account for the internal structure of incoming nucleon and for 
the possibility of nucleon excitations, $p\to N^*$, in the intermediate states.

On the other hand, the model~\cite{1312}, which we use, corresponds to a two-channel eikonal. That is the nucleon wave function is described by a superposition of two Good-Walker~\cite{GW} (GW) diffractive eigenstates. These are eigenstates with respect to the high energy (Pomeron exchange) interaction\footnote{There are no transitions between the different eigenstates caused by the Pomeron.}. So to implement the two-channel eikonal we have to repeat the prescription described above for each combination of the GW  eigenstates
using the corresponding opacities $\Omega_{ij}$ where the indices $i,j=1,2$ denote the GW state in the fast (beam) and target nucleon respectively.

\subsection{Effect of the enhanced diagrams}
The correction caused  by  `enhanced' screening can be calculated using the AGK reggeon cutting rules~\cite{AGK}. These rules relate the cross section of high-mass diffractive dissociation with the value of absorptive correction. 

 Now  diffractive dissociation plays the role of elastic cross section which was used to fit the eikonal proton opacity $\Omega(b)$ while calculating the $S^2_{\rm eik}$ survival factor (\ref{screen}). The difficulty is that, unlike the elastic cross section, the experimental data on high-mass dissociation are quite scarce. Based on the preliminary TOTEM data~\cite{TOTEM3} we assume that the cross section of single 
dissociation (integrated over the $M_X=3.4~-~1100$ GeV mass interval) is $\sigma^{\rm SD}=6.5$ mb 
\footnote{ The small value, 6.5 mb, is explained by the smallness of triple-Pomeron vertex and strong eikonal absorption.}
 and that the mean $t$-slope\footnote{In the three measured
$M_X$ mass intervals the values of slope were found to be $B_{\rm dis}=10.1,\ 8.5,\ 6.8$ GeV$^{-2}$~\cite{TOTEM3}.} is $B_{\rm dis}=8.5$ GeV$^{-2}$.

Next, in order to estimate the contribution of the diagrams\footnote{The significant role of these diagrams was emphasized in~\cite{Kop2}.} of the
 type of Fig.~\ref{fig:6}(b)
 we use the Pomeron piece of the pion-nucleon
 cross section 
 \be
 \sigma(\pi N)=13.63~(s_{\pi N}/1~\mbox{GeV}^2)^{0.0808} ~~{\rm mb}~,
 \ee
  in
the Donnachie-Landshoff~\cite{DL} parametrization, and the slope $B_{\pi N}=6$ GeV$^{-2}$ (see e.g. Fig.10 of~\cite{Dino}). Strictly speaking both the cross section and the slope depend on $x_L$ and the transverse momentum of the neutron. Here we adopt representative values since the result is comparatively insensitive to the exact numbers. Indeed, due to the relatively small $t$-slope (in comparison with that of the elastic scattering), the
corresponding screening amplitude comes from the
region of large impact parameters, $b_t$. The low $b_t$ domain is already strongly suppressed by eikonal absorption (\ref{screen}), while at  larger $b_t$ the `tail' of the remaining enhanced screening amplitude is rather small. Therefore this component of screening only weakly affects the final result. The same is valid for $\pi N$ absorption, Fig.~\ref{fig:6}(b).
Thus it is sufficient to calculate the effective $\pi N$ and `enhanced'  opacities, $\Omega_{\rm enh}$, in a simplified way as
\be
\Omega_{\pi N}(b)=
\frac{\sigma(\pi N)}{2\pi B_{\pi N}}e^{-b^2/2B_{\pi N}}\ ,
\ee  
\be
\label{enh-op}
\Omega_{\rm enh}(b)=\frac{\sigma^{\rm enh}}{2\pi B_{\rm dis}}e^{-b^2/2B_{\rm dis}}\ ,
\ee
where the effective cross section $\sigma^{\rm enh}=14.1$ mb was recalculated\footnote{We do not include here the cross section of low-mass dissociation, since in the case of a two-channel eikonal the low-mass dissociation is reproduced by the non-zero  dispersion of the individual GW component cross sections $\sigma_{ij}$  (see~\cite{1312} for the details). We take
only a half of the whole $\sigma^{\rm SD}$ since the experimental number accounts for the dissociation of {\em both} protons, while here we have to consider high-mass dissociation of the target proton only.}  based on the TOTEM data as 
\be
\sigma^{\rm enh}=(\sigma^{\rm SD}/2)B_{\rm dis}/(\sigma^{\rm tot}/16\pi).
\ee

Combining these results together, the opacity 
 $\Omega(\vec b-\vec b_1)$ in (\ref{Ib}) is replaced by the sum
\be
\Omega=\Omega_{{\rm eik},ij}(\vec b-\vec b_1)+\Omega_{\pi N}(b)+
\Omega_{\rm enh}(b)
\label{Omega1}
\ee
with $\Omega_{{\rm eik},ij}$ corresponding to the opacity in the 
interaction of the $i$ and $j$ GW components.

Let us examine this last modification in more detail.
In fact, not all inelastic interactions populate the rapidity gap and reduce the neutron energy fraction, $x_L$. 
Part of the inelastic events have, from the beginning, no secondaries within the gap interval.
First, there are events with dissociation of the target proton. It is evident that for target proton dissociation no new secondary particles are produced within the rapidity gap interval between the fast neutron and the remaining system $X$. 
 Next, with 
some probability, $P(x_L)$, such a (moderately large) gap could be formed at the hadronization stage \cite{Khoze:2010by}. Assuming that, in a standard inelastic event, the neutron distribution is
\be
\frac{dN}{dx_n}\simeq {\rm const}
\ee
we get $P(x_L)=1-x_L.$
Therefore we have to multiply the  full opacity by $1-P(x_L)$ and in addition multiply the `eikonal' opacity $\Omega_{{\rm eik},ij}$ by the factor 
\be
1-\sigma^{\rm SD}/2\sigma_{\rm inel}=1-6.5/2(98.7-24.9)=0.956
\ee
 to account for proton dissociation. 
So, finally, (\ref{Omega1}) is altered so that the full $\Omega$ in (\ref{Ib}) becomes
\be
\Omega=x_L~\left(~0.956~\Omega_{{\rm eik},ij}(\vec b-\vec b_1)+
\Omega_{\pi N}(b)+\Omega_{\rm enh}(b)\right)\ .
\ee

The absorptive factors for the leading $\Delta$-isobar production, and for the $\rho$ and $\ a_2$ exchange amplitudes are calculated in a similar way.

\section{Isolation of $\pi$ exchange in leading neutron LHCf data  \label{sec:4}}
We have seen that the inclusive leading neutron cross section is  not {\em totally} given 
by the simple pion-exchange formula (\ref{pi-reg}). Above we have studied several other effects. We have enumerated contributions from $\rho$ and $a_2$ exchanges, and from neutrons coming from the
 $\Delta$-isobar or from diffractive nucleon excitations decays, $N^*\to n\pi$. Next, we discussed absorptive corrections: indeed, we considered both eikonal and enhanced screening effects.  So in order to confront the LHCf data on forward neutrons we should explore the kinematic domains of the data where (a) $\pi$ exchange  dominates and (b) the original `Born' amplitude is minimally modified.

\subsection{Form factor of the $\pi N$ vertex}
 Even for pion exchange the form factor of the pion-nucleon vertex is poorly known. This is not a big problem when we are working close to the pion pole, say, using the LHCf data for neutron rapidities $\eta>10.76$ and looking for the neutrons with $E_n=3.25-2.5$ TeV which correspond to a mean $-t=0.02-0.08$ GeV$^{2}$ respectively.
In this case a variation of the slope of the form factor, $F(t)=e^{Bt}$, by $\delta B=\pm 1$ GeV$^{-2}$ will lead to a $2\delta B(m^2_\pi-t)\sim \pm (8-20)$\%
variation of the result respectively. Already at this stage we see that at lower $x_L$ the theoretical uncertainty increases and it is safer not to go below $x_L=0.75$ (that is, the $E_n=2.75-2.5$ TeV bin). The situation becomes much worse for a smaller rapidities. In particular, for the case of $\eta=8.99-9.22$  the mean $|t|$ is about 0.5 GeV$^{2}$ leading to up to a factor of 0.4 to 2.7 uncertainty.\footnote{Besides this, at  larger $q_t$ values, the relative contribution of
the $\rho,\ a_2$ trajectories increases since the corresponding vertices have a very large spin-flip component
which is proportional to $q_t$.}
Therefore below we  consider only the largest $\eta>10.76$ rapidity interval. 

Recall that in our calculation  we used non-reggeized pion exchange. If  
 instead, the reggeized version of (\ref{pi-reg}) was implemented with a
 vertex form factor $F(t)={\rm exp}(1.5(t-m^2_\pi))\;$  (where $(t-m^2_\pi)$ is 
in GeV$^2$) 
the results change only by $\pm 2$\%  (where +2\% is for the $E_n=2.75-2.5$ TeV bin).

\subsection{Screening effects  \label{sec:4.2}} 
Besides this, for larger $|t|$, the screening effects become stronger.
 For the $\eta=8.99-9.22$ interval the full survival factor is rather small, namely $S^2=\langle e^{-\Omega} \rangle \simeq 0.032-0.075$; that is $\langle \Omega \rangle \sim 3$. So due to the exponential dependence  even a moderate theoretical uncertainty in the calculation of $\Omega$ could strongly influence the result.
For larger rapidities $\eta>10.76$ and $x_L>0.75$  the major contribution comes from relatively large impact parameters where the nucleon is not so black; that is where the 
optical opacity $\Omega$ is not large. Here, for $E_n=3.25-2.5$ TeV, the mean survival factor is respectively $S^2=0.45-0.3$ and  within an accuracy of (5-10)\% we can rely on the calculation of
the absorptive corrections. 
Indeed, using, instead of the two-channel eikonal model~\cite{1312}, a one-channel approach with the opacity 
taken  just from the experimental data multiplied by the `semi-enhanced' factor $C=1.3$~\cite{oldKaid} to account for possible $N^*$ intermediate states (and neglecting the Re/Im ratio\footnote{The real part of the elastic amplitude, was accounted for in our calculations of the rescattering corrections. It enlarges the final cross section by less than 1\%.}) we obtain a cross section larger by about 6 - 12\% only.

Neglecting completely enhanced screening enlarges the cross section by about 10-20\%, while replacing the Donnachie-Landshoff Pomeron contribution to the pion-nucleon cross section by $\sigma_{\pi N}=26$ mb we obtain a result smaller by 4 -8\%. These numbers correspond to the  $E_n=3.25-2.5$ TeV interval.

\subsection{$\rho, a_2$ and $\Delta$ effects}
The contribution coming from the secondary $\rho$ and $\ a_2$ reggeons calculated using the couplings obtained in~\cite{Irving} based on the Vector Meson Dominance model   
is rather large in the highest $x_L$ bin (with $E_n$ in the $3.5-3.25$ TeV bin and $\eta>10.76$). Assuming the equal meson-proton cross sections ($\sigma(\rho p)=\sigma(a_2 p)=\sigma(\pi p)$), it amounts to
 37\% of the pion exchange term.
However, in 3 bins with lower $E_n$ the contribution of the $\rho$ and $\ a_2$ diagrams decreases to (12 - 9)\%.
Bearing in mind  large experimental error ($46$\%) and the large  admixture of
the $\rho$ and $\ a_2$ exchange processes in the highest $x_L$ bin, we prefer not to use this
kinematic region for extracting the high energy pion-proton cross section. 

The contribution from the $\Delta$-isobar decay in this domain is  practically negligible. Calculating the cross section of $\Delta$ production using the couplings from~\cite{Irving}, after the decay we get less than a 1.1 - 2.5\% correction.

\subsection{Migration}
 The next problem is migration. After an additional soft interaction the fast nucleon
may change its momentum and  `migrate' from one kinematical bin to another.  This possibility was considered in detail in~\cite{Kaidalov:2006cw} where it was shown that for low $q_t<0.1$ GeV  migration practically does not affect the neutron spectra at $x_L>0.75$, and thus could be neglected in the region of 
interest.\footnote{One rescattering gives about 1.6$-$7.1\% contribution in the $E_n=3.25-2.5$ region.
Note that after the $S^2$ absorption is taken into account, only the large $b_t$ contributions survive, and the mean number of rescatterings, $\langle\nu\rangle=\langle\Omega\rangle$, is less than 0.3 for the  $E_n=3.25-3$ TeV bin,
and less than 0.6 for the $E_n=2.75-2.5$ TeV bin.}

\subsection{Low-mass diffractive proton excitations   \label{sec:4.5}}
A more serious problem arises from neutrons produced in the decay of low-mass diffractive proton excitations, $N^*\to n \pi$. At $\sqrt s=7$ 
TeV, the TOTEM result~\cite{TOTEM1} for the cross section of low-mass proton 
 dissociation is $\sigma^{\rm D}_{lowM}=2.6\pm 2.2$ mb, with $M_X<3.4$ GeV; this measurement corresponds to allowing both protons to diffract.  That is,
the cross section of {\em one} proton dissociation is about 1.5 mb. Note that part of this cross section is already included in the pion-exchange contribution. Indeed, keeping the elastic component in the total pion-proton cross section, we include the $pp\to (n+\pi^+)+p$ process where in almost the whole essential kinematic region the mass of the $n\pi$ system is less than 3.4 GeV. Accounting for the screening corrections, this Drell-Hiida-Deck~\cite{Deck} contribution is equal to
\be
\sigma^{\rm DHD}=0.026~\sigma_{\rm el}(\pi p)\sim 0.2-0.3 ~~ {\rm mb}.
\ee
Thus we still have more than 1 mb of diffractive proton dissociation which, in its decay, could produce leading neutrons. Unfortunately there is insufficient information at the LHC energies. We do not know the $M_X$ mass distribution, the $t$-slope of the low-mass dissociation, and the possible polarization of the $N^*$ resonances. Looking at the lower energy data we assume that the dominant contribution comes from the region of $M_X\sim 1.7$ TeV and that the $N^*$ system is produced with the same slope as that in  elastic $pp$-scattering; that is $B_{\rm dis}=20$ GeV$^{-2}$~\cite{TOTEM2}.

 At large values of the neutron $x_L>0.75$ the  main contribution arises 
from the two-body $N^*\to n\pi^+$ decay. For higher multiplicity it becomes 
difficult to allow for such a large neutron momentum fraction.
 We assume a non-polarized decay with the 
branching ratio\footnote{The $N^* \to n\pi^+ $branching ratio Br $\simeq$ 1/3 comes from about 50$\%$ $N^* \to N\pi$ 
branching, with the other 
50$\%$ due to the $N\pi\pi$ and $\Delta\pi$ decay channels (these are the typical branching ratios
for $N^*$ resonances in the 1400 - 1700 MeV region\cite{PDG}). Finally a factor 2/3 comes from 
the isotopic spin factor
Br$(p^* \to n\pi^+)/$Br$(p^*\to p\pi^0)=2$. Note that the resulting cross section 
$\sigma(N^*\to n\pi^+)\simeq$ 0.33 mb
is in agreement with the 
lower energy data~\cite{babaev} ($\sigma\sim 0.3$ mb) assuming that the flat energy 
dependence continues up to LHC energies. } Br $\simeq$ 1/3. 

 The cross sections that we find,  assuming $\sigma(N^*)=1$ mb, can be rather large, see the 3rd column of Table \ref{tab:diff}. In the highest $E_n=3.5-3.25$ TeV bin they could account for up to 25\% of the leading neutron cross section. For the next three bins this contribution becomes negligible in comparison with the experimental error bars of the LHCf data.

\begin{table}[hbt]
\label{tab:diff}
\begin{center}
\begin{tabular}{|c|c|c||c|c|c||c|c|}\hline
$E_n$ (TeV)& LHCf data & $\Delta\sigma^{\rm diff}$& $\pi$&$\rho+a_2$&$\Delta$&migr&$\langle 1-x_L\rangle$ \\
\hline
3.5$-$3.25&$232\pm 106$& $58$&2.41&0.87&0.01&0.2&0.047\\
  3.25$-$3 &$249\pm 78$& 9.6&5.62&0.66& 0.06&0.6&0.109\\
  3$-$2.75 &$282\pm 48$& 1.6&5.53&0.50&0.09&1.7&0.177\\
2.75$-$2.5 &$298\pm 34$& 0.4&3.75&0.34&0.09&5&0.247\\
 \hline
\end{tabular}
\caption{\sf   The 2nd and 3rd columns show, respectively, the cross sections ($\mu$b/TeV) for leading neutrons as measured by LHCf \cite{LHCf} and the contribution $\Delta\sigma^{\rm diff}$ coming from the decay of low-mass  proton excitations, $N^*\to n\pi$, calculated as described in Section \ref{sec:4.5}. The $i=\pi,\rho+a_2,\Delta$ columns are the ratios $R^i=(d\sigma^i/dE_n)/\sigma(\pi^+ p)$ of the calculated inclusive cross section to the total pion-proton cross section. The ratios presented here are measured in inverse TeV and multiplied by a factor of 1000, so that the LHCf result divided by $R^i$ gives the value of $\sigma(\pi^+ p)$ in mb;
for example, for the 3$-$2.75 TeV bin, accounting for $\pi$-exchange {\it only} we obtain $\sigma(\pi^+p)= 282/5.53 \simeq51$ mb; since the $\pi$ fraction is 88$\%$ (see Table \ref{tab:sig}) this results in the true $\sigma_{\rm tot}(\pi^+p)\simeq45$ mb.
 Finally, the `migr' column shows, as a \%, the effect of fast neutron rescattering (that is, migration of the leading neutron). The last column shows the mean value of the momentum fraction carried by the pion in the case of the pion exchange contribution.}
\label{tab:diff}
\end{center}
\end{table}
\begin{table}[h]
\begin{center}
\begin{tabular}{|c|c|c|c||c|c||c|c|}\hline
$E_n$ (TeV)& LHCf data & $\sigma_{\rm tot}(\pi p)$ &$\sqrt{s_{\pi p}}$ (TeV)& $\sigma^{\rm Reg}$&$\sigma^{\rm  Comp}$& $\pi$ fraction& $S^2_\pi$ \\
\hline
3.5$-$3.25 &$232\pm 106$& $52.7\pm 32.1$&1.52&44.6&60.1&0.55&0.56\\
3.25$-$3 &$249\pm 78 $& ${\bf 37.5\pm 12.2}$&{\bf 2.31}&47.7&65.9&0.85&0.44\\
3$-$2.75 &$282\pm 48 $& ${\bf 45.0\pm 7.7}$&{\bf 2.94}&49.6&69.4&0.88&0.36\\
2.75$-$2.5 &$298\pm 34 $& ${\bf 67.9\pm 7.7}$&{\bf 3.48}&50.9&71.9&0.85&0.32\\
 \hline
\end{tabular}
\caption{\sf The 3rd column is the $\pi^+ p$ total cross section (mb) extracted from the LHCf leading neutron data $d\sigma/dE_n$ ($\mu$b/TeV) shown in the 2nd column. The result for the first $E_n$ bin is not reliable (see
 the huge error bar), and is shown only for completeness. The 4th column is the mean pion-proton energy corresponding to the particular $E_n$ bin. The value of the $\pi^+p$ cross section (mb) obtained from the
extrapolation of a simple Regge pole fit \cite{DL} and from the Compete fit \cite{RPP} to lower energy hadron-hadron cross section data are shown for comparison in the 5th and 6th columns respectively.  The last two columns show the relative contribution of the pion exchange process to the total leading neutron cross section and the pion exchange gap survival factor respectively.}
\label{tab:sig}
\end{center}
\end{table}

\section{$\pi^+ p$ cross section from LHCf leading neutron data}
\begin{figure}
\begin{center}
\vspace*{-5cm}
\includegraphics[height=15cm]{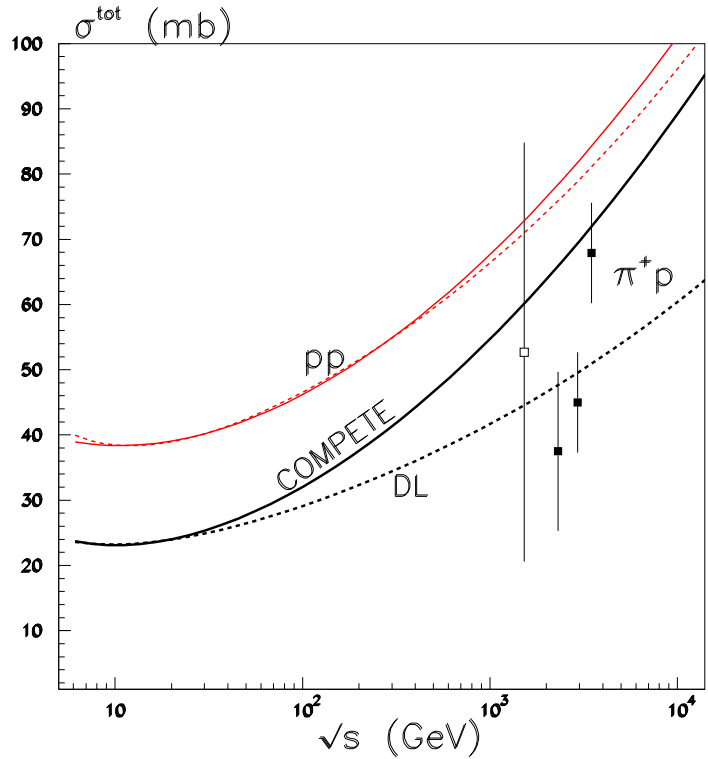}
\caption{\sf The four values of $\pi^+ p$ total cross section that we extract from the LHCf data on leading neutrons \cite{LHCf}, compared with expectations based on fits
 to lower-energy hadron-hadron total cross section data parametrized by two Regge poles, DL \cite{DL}, or using the COMPETE parametrization \cite{RPP}. Note that the results of both parametrizations coincide in the region of the existing $\pi p$ cross section data, that is for $\sqrt{s}<25$ GeV. For reference, the upper two (red) curves are the corresponding descriptions of the $pp$ total cross section. Recall that
 the error bars shown here reflect the experimental uncertainties {\em only}. 
 The possible theoretical uncertainties are discussed in detail in the main body of 
 paper, see Sections 2~-~4.
\label{fig:x}}
\end{center}
\end{figure}
We use the different contributions to forward neutron production in $pp$ collisions described above, together with LHCf data \cite{LHCf}, to extract the $\pi^+ p$ total cross section at various energies $\sqrt{s_{\pi p}}$.
The results are shown in Table \ref{tab:sig}.
We show only the errors coming from the experimental error bars. The uncertainties arising from the theoretical approach were discussed in Section \ref{sec:4}, and the sizes of the individual contributions are shown in the central part of Table \ref{tab:diff}. As expected from Section \ref{sec:4}, we see that the result for the highest $E_n$ bin (that is, the bin with the highest $x_L$, which corresponds to the lowest pion-proton energy $\sqrt{s_{\pi p}}$) is not reliable, and is shown only for completeness.  On the other hand, we expect better theoretical accuracy for the next three experimental $E_n$ bins where we 
 have a larger fraction of $\pi$ exchange.
It is clearly seen from Table \ref{tab:sig} and Fig.~\ref{fig:x} that the pion-proton cross section increases with energy, however the uncertainties are rather large.

Fig.~\ref{fig:x} compares the values of $\sigma_{\rm tot}(\pi p)$ extracted from the LHCf data with two predictions based on extrapolations of fits to lower energy hadron-hadron cross sections, shown by the lower two curves labelled DL \cite{DL} and COMPETE \cite{RPP}.  The large error bars do not allow us to decide between the two extrapolations.  For reference we also show, by the upper two curves, the DL and COMPETE descriptions of the total $pp$ cross section.

The values of the $\pi^+ p$ cross section that we obtain are smaller than those of \cite{Ryutin:2016hyi} extracted from the same LHCf data but at a lower rapidity interval $8.99<\eta<9.22$. Recall, however, that at lower rapidities we deal with relatively large $q_t\sim 0.6$ GeV, that is with $|t|\sim 0.4$
 GeV$^2$, where the uncertainty in form factor can appreciably change the result. Moreover, nothing is said in~\cite{Ryutin:2016hyi} about the effects of migration, proton diffractive dissociation and the enhanced absorptive corrections.
 The role of all these effects was described in Section 4 above and, since for $\eta>10.76$ we work much closer to the $\pi$-pole, we believe our results, shown in Table \ref{tab:sig} and Fig.~\ref{fig:x}, are more reliable.
 
 Nevertheless, one has to remember that the extraction of the pion-proton cross section from leading neutron inclusive data is not so straightforward.
 To describe the full `kitchen' of effects hidden in this procedure is one of the goals of our paper.

\begin{figure}
\begin{center}
\vspace*{-5cm}
\includegraphics[height=15cm]{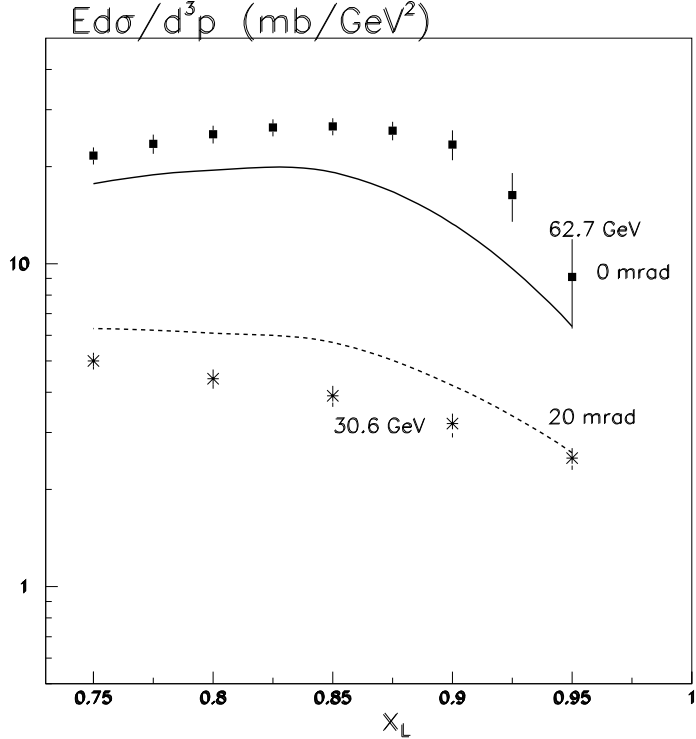}
\caption{\sf  The description of the CERN-ISR leading neutron data \cite{data1,data2}
\label{fig:ISR}}
\end{center}
\end{figure}
\section{Description of the CERN-ISR data}
In order to check the quality of our approach, in Fig. \ref{fig:ISR} we use the same formalism (as that we used to describe the LHCf data) to calculate the leading neutron cross sections measured in the CERN-ISR energy range
 for $\sqrt s_{pp}=30.6-62.7$ GeV \cite{data1,data2}. 
The description of the data of the two experiments is puzzling.
 We underestimate the data obtained at zero angle ($q_t=0$), but overestimate the data obtained at 20 mrad. Note however that the two groups of data comes from different experiments and reveal some inconsistency. It is hard to provide the steep $q_t$ dependence that is needed to reconcile both data sets\footnote{The energy dependence in each experiment was rather weak and the data are consistent with the scaling behaviour; that is at a fixed $q_t$ and $x_L$ the cross section does not depend on $\sqrt s$.} with reasonable slopes of the vertex form factors $F(t)$.

Moreover, contrary to the zero degree case, the 20 mrad curve in Fig.~\ref{fig:ISR} is the {\em minimal} prediction. It includes only the $\pi,\rho$ and $a_2$ contributions and neglects the $p\to N^*\to n+X$ dissociation which in some papers (e.g.~\cite{babaev,Ponom}) was described completely via the Deck process~\cite{Deck}~\footnote{However, without accounting for the gap survival factor $S^2$.}.
 
It was suggested by
Kopeliovich et al. \cite{Kop2} that most probably the data
at $q_t=0$ have unreliable normalization. On the other hand Kaidalov's group
trust more the zero angle data; recall that in~\cite{data2} it was emphasized that in the first experiment~ \cite{data1} the background was rather high.
Our prediction is somewhere in-between the two data sets. Recall that, in comparison with the LHCf data, the CERN-ISR data are at much lower energies, where the secondary Reggeon contributions are not negligible and other effects not discussed here may be present; nevertheless the accuracy of our description should still be reasonable.

 \section{Conclusion}
We discuss the different contributions to the leading neutron inclusive spectra of LHCf \cite{LHCf}.
Besides pion exchange, as $x_L\to 1$ an important role is played by the
$\rho$ and $a_2$ trajectories. In addition we have to account for the neutrons
coming from diffractive dissociation such as $pp\to (n\pi^+) + p$ and for the final state rescattering of the leading baryon, which leads to 'migration' of the leading neutron from one to another kinematical bin.
Nevertheless there exists a {\it small} kinematic domain ($x_L\sim 0.75-0.9$ and
$q_t<m_\pi$) where the pion pole dominates\footnote{Essentially only in 3 bins of over the 40 bins of data collected by LHCf \cite{LHCf}} and the $\pi$-exchange amplitude provides
more than 80\% of observed cross section. The data
collected in three bins in this region can be used to extract the value of the $\pi^+ p$ total
cross section, see Table \ref{tab:sig} and Fig.~\ref{fig:x}.

Recall that even here we have to account for the absorptive corrections (that is, include a gap
survival factor $S^2$) which suppresses the original (Born)
cross section by more than a factor of two (see the last column of Table \ref{tab:sig}}). However, in this {\it small} $q_t$ region the value of $S^2$ can be {\it reliably} calculated with good accuracy based on the data for elastic $pp$-scattering which allow a good
determination of the proton optical density (that is, the opacity, $\Omega(b)$).
Of course there is some uncertainty depending on the particular model used to
describe the differential elastic cross section, but as we demonstrated in Section \ref{sec:4.2}, this uncertainty is not too large.

Actually the main aim of our paper is not just to extract the pion-proton cross
section, but
rather to explain all the subtleties
 hidden in the procedure in order to give an understanding of
the possible theoretical uncertainties.  One outcome is that it is indeed possible to find a kinematic region where the pion-pole dominates. However, even in this case it is critical to account for the $S^2$ absorptive correction, which, as mentioned above, appreciably affects the value of the cross section.

Within the experimental error bars the results obtained for $\sigma_{\rm tot}(\pi p)$ are consistent with the
extrapolation given by Donnachie-Landshoff~\cite{DL} or COMPETE~\cite{RPP}
parametrizations.  The present indications are that the $\pi p$ cross section
rises with energy steeper than  in the proton-proton case.

 \section{Outlook}
 
 The present leading neutron data, and hence our determination of $\sigma_{\rm tot}(\pi^+ p)$ 
 in the few TeV energy region, are not yet sufficiently accurate to be very informative. But
 as the experimental statistics improve, it should be possible, with the framework we 
 discussed, to make a good determination of the high energy dependence 
 of $\sigma_{\rm tot}(\pi^+ p)$.  Moreover, when the 13 TeV data become available, it will be 
 possible to extend the energy reach of the measurements and to enter the region which can 
 distinguish between the
extrapolations (for example \cite{DL,RPP}) from lower energies.

To obtain a more precise result and to better fix the parameters it would be
valuable to measure the $q_t$ dependence of the leading neutron spectra. As discussed in
\cite{Sobol:2010mu},
this could be achieved in a CMS measurement with the Zero Degree
Calorimeter.
Engaging the Forward Shower counters (FSC) \cite{Albrow:2008az} would allow  the suppression of the contribution
arising from low-mass dissociation of the beam proton.

On the other hand, in the common runs of LHCf with ATLAS,
 it will be possible to study
the low-mass diffractive proton dissociation, $p\to N^*\to n+X$, contribution
and to exclude this component
from the inclusive (non-diffractive) neutron cross section. Again, FSC analogous to \cite{Albrow:2008az} will allow a better selection of low-mass dissociation.

Moreover, ATLAS could measure the
{\it distribution of secondaries} in the events containing  a leading neutron. In this way
we have a chance to study not only the value of $\sigma_{\rm tot}(\pi^+ p)$ but also
the inclusive
cross sections in the $\pi^+ p$ collisions as well.

\section*{Acknowledgement}
We are grateful to Takashi Sako and Sergey Ostapchenko for useful discussions. 
MGR thanks the IPPP at Durham University for hospitality and
 VAK thanks the Leverhulme Trust for an Emeritus Fellowship.
 The research of MGR was supported by the RSCF grant 14-22-00281.

\end{document}